\documentclass{elsarticle}

\usepackage{lineno,hyperref}
\modulolinenumbers[5]

\journal{Journal of Theoretical Biology}

\usepackage{url}
\biboptions{authoryear}

\usepackage{amsfonts}
\usepackage{amsmath,amsthm,amssymb}
\usepackage{graphicx}
\usepackage{caption}
\usepackage{subcaption}
\usepackage{hyperref}

\newcommand{\del}{\partial}

\begin{document}

\begin{frontmatter}

\title{Transport Equations for Subdiffusion with Nonlinear Particle Interaction
}


\author[mymainaddress]{P. Straka\corref{mycorrespondingauthor}}
\cortext[mycorrespondingauthor]{Corresponding author}
\ead{p.straka@unsw.edu.au}
\author[sergeiaddress]{S. Fedotov}
\ead{sergei.fedotov@manchester.ac.uk}

\address[mymainaddress]{School of Mathematics and Statistics, UNSW Australia, Sydney NSW 2052}
\address[sergeiaddress]{School of Mathematics, The University of Manchester, Manchester M13 9PL, UK}

\begin{abstract}
We show how the nonlinear interaction effects `volume filling' and `adhesion' can be incorporated into the fractional subdiffusive transport of cells and individual organisms. 
To this end, we use microscopic random walk models with anomalous trapping and systematically derive generic
\textit{non-Markovian and nonlinear} governing equations for the mean concentrations of the subdiffusive cells or organisms.
We uncover an interesting interaction between the nonlinearities and the non-Markovian nature of the transport.
In the subdiffusive case, this interaction manifests itself in a nontrivial combination of nonlinear terms with fractional derivatives.
In the long time limit, however, these equations simplify to a form without fractional operators.
This provides an easy method for the study of aggregation phenomena. 
In particular, this enables us to show that volume filling can prevent ``anomalous aggregation,'' which occurs in subdiffusive systems with a spatially varying anomalous exponent.

\end{abstract}


\end{frontmatter}

\linenumbers

\section{Introduction}

Stochastic models for the diffusive motion of biological cells and organisms  
are well established in the mathematical biology community. Random
walk models, stochastic differential equations and their governing nonlinear partial
differential equations have been very successful from a mathematical
modelling standpoint. They provide tractable means to incorporate various
taxis effects such as the directed transport along the concentration
gradient of external signals \citep{Othmer2002a,Hillen2009,Stevens00},
particle generation and degradation rates which depend on particle
concentrations \citep{Murray2002,Oelschlager1989}, density dependent
dispersal rates \citep{Mendez2012,Murray2002}, volume exclusion effects
\citep{Painter2002,Simpson2011,Fernando2010}, and adhesion between particles
\citep{Anguige2011,Armstrong2006,Johnston2012}. A defining feature of most such
nonlinear reaction-diffusion-taxis equations is that the macroscopic
transport processes involving diffusion and advection are derived from
microscopic Markovian random walk models;
see the excellent review by \cite{Stevens1997a}.
However, this does not fit well
with anomalous non-Markovian subdiffusive systems, for which the transport
operators are non-local in time and the mean squared displacement of
individual particles grows proportionally to $t^{\mu }$, where $0<\mu <1$ \citep{Metzler2000}.
Anomalous transport occurs microscopically on the level of individual cells,
e.g.\ for the transport of macromolecules within living cells \citep{Golding2006,TMT04,Weiss2004,Banks2005}.
Moreover, it has been found that the motion of individual cells is anomalously diffusive
\citep{Dieterich2008,mierke2011integrin,Fedotov2013a}.

The main mathematical models for subdiffusive dynamics are the Continuous
Time Random Walk (CTRW) and fractional Brownian motion (fBm). Both processes
are non-Markovian, unlike Brownian motion. The CTRW appears to be the most
popular model for anomalous dynamics \citep{Metzler2000}, presumably because
it admits a tractable PDE formalism \citep{BMK00,HLS10PRL}. However, it
should be noted that most articles on anomalous transport deal with linear
fractional PDEs without particle interactions. Unlike for Markovian
dynamics, it is challenging to incorporate nonlinearities into the
subdiffusive PDEs. For instance, even if the particle death rate is bounded
below, by naively adding a degradation term to the PDE one can achieve
negative particle concentrations \citep{Henry2006}.
Transport equations for CTRWs with nonlinear reactions have only recently
been derived \citep{Mendez2010,Angstmann2013}. Apart from an article by one of the authors \citep{Fedotov2013b},
to our knowledge, particle interactions have not yet been
incorporated into the CTRW framework. The challenge is to take into account
non-linear effects: volume exclusion \citep{Painter2002} and adhesion \citep{Anguige2011} together with  subdiffusive transport.

The main purpose of this article is to systematically derive generic
\textit{non-Markovian and non-linear} integro-differential equations for the mean
concentration of particles such as randomly moving cells or individual organisms.
Our aims are: (i) to understand the interaction of non-Markovian transport and nonlinearities due
to volume filling and adhesion effects, and
(ii) to find the stationary solutions of nonlinear non-Markovian transport equations that describe aggregation phenomena.

On our way towards goal (i), we give a formalism which connects nonlinearly interacting microscopic CTRWs with nonlinear and non-Markovian diffusion equations. 
As it turns out, our formalism also applies to the situation where the anomalous exponent $\mu$, which governs the trapping behaviour of the CTRW, varies in space \cite{Chechkin2005a}.
This situation is very significant for biology because
it may explain the widespread phenomenon of anomalous accumulation of
bacteria in particular patches. 
One example is the aggregation of phagotrophic
protists \citep{fenchel1999motile}, where ``cells become immobile in attractive patches,
which will then eventually trap all cells.'' 
Another example is the
formation of nodules on the roots of nitrogen-fixing plants that contain the
colony of nitrogen-fixing bacteria 
\citep{wadhams2004making}. 

It is well known that the movement of bacteria in environments with varying favorability is in the most cases determined by chemokinesis rather than chemotaxis.
The reason for this is that typically the bacteria/cells are too small to sense the macroscopic gradient of a chemotactic substance $S(x)$ \citep{erban2005signal}.  
Hence a model for the random motility of microorganisms should take into account the dependence of the transition probability $\gamma$ on the nonuniformly distributed concentration $S(x)$, 
rather than the dependence of a cell's jump direction on the gradient $\partial S / \partial x$. 
With this in mind, CTRWs with space-varying anomalous exponent $\mu$ arise very naturally as models for chemokinesis: 
Suppose that $\mu = \mu(S(x))$ is a decreasing function of a favourable substance with concentration $S(x)$. 
Then the transition probability $\gamma$ (i.e.\ the probability of a jump away from $x$) equals 
\begin{equation*}
\gamma \left( \tau ,S\left( x\right) \right) =\frac{\mu ( S\left(
x\right) ) }{\tau _{0}+\tau },
\end{equation*}
where $\tau$ is the residence time and $\tau_0$ is a constant (see Eq.\eqref{eq:chemokinesis}). 
Hence the rate at which a bacterium jumps away from a favourable environment at $x$ is small, 
which leads to the phenomenon of anomalous aggregation \cite{Fedotov2012}.

The setup is as follows: In Section \ref{sec:markov} we quickly reiterate the derivation of nonlinear Markovian transport equations from microscopic stochastic models. Section \ref{sec:non-markov} contains a quick overview over the anomalous
sub-diffusion literature and fractional diffusion equations. In Section \ref{sec:nonlin-nonmarkov} we use the structured density approach and recover
Markovian methods for CTRWs; this allows for the derivation of nonlinear
differential equations involving subdiffusion.
Finally, in Section \ref{sec:profiles} we give examples of stationary solutions to nonlinear
fractional PDEs that describe the aggregation phenomenon.

\section{Markovian transport with nonlinear particle interaction}

\label{sec:markov}

In this section, we briefly review the standard derivation of nonlinear diffusion equations, starting from a microscopic random walk model.
For simplicity, we consider a one dimensional lattice of sites $x$ which are
evenly spaced with spacing $h$. We study the dynamics of the concentration $%
\rho (x,t)$ of particles (e.g.\ cells, bacteria, etc.). We assume that
particles perform instantaneous jumps to neighbouring lattice sites. We
write $T^{+}(x,t)$ and $T^{-}(x,t)$ for the rates of jumps to the right
resp.\ left. Rates are instantaneous and may vary in space $x$ and in time $%
t$. The total jump rate is then $T(x,t):=T^{+}(x,t)+T^{-}(x,t)$. The master
equation for $\rho (x,t)$ reads
\begin{equation}
\frac{\partial \rho (x,t)}{\partial t}=T^{+}(x-h,t)\rho
(x-h,t)+T^{-}(x+h,t)\rho (x+h,t)-T(x,t)\rho (x,t).  \label{eq:markov-master}
\end{equation}%
Transport models for diffusion, chemotaxis, volume filling and adhesion have
been studied by \cite{Anguige2011}, \cite{Anguige2009} and \cite{Painter2002}.
A general model which accommodates all the above effects is given by
\begin{equation} \label{eq:Tpm-def}
T^{\pm }(x,t)=\lambda _{0}\left( 1-[S(x\pm h,t)-S(x,t)]\right) q(\rho (x\pm
h,t))a(\rho (x\mp h,t))
\end{equation}%
Here, $\lambda _{0}$ is the rate parameter, and 
$S(x,t)$ is a spatio-temporally varying external signal
(e.g.\ a chemoattractant or chemorepellent concentration).
The functions $q(\rho )$ and $a(\rho )$ model volume filling and adhesion phenomena;
they are decreasing with respect to the concentration density $\rho (x,t)$ and map
to values in $[0,1]$. The
volume filling function $q(\rho (x,t))$ can be interpreted as the
probability that a particle will be accommodated at $x$, should it attempt
to jump there at time $t$. With the remaining probability $1-q(\rho (x,t))$,
it will not find enough room at $x$ and hence will not jump. Similarly, the
adhesive effect is modelled with the function $a(\rho (x,t))$: Given that a
particle attempts to jump from $x$ to $x+h$ at time $t$, it succeeds in
jumping there with probability $a(\rho (x-h,t))$. With probability $1-a(\rho
(x-h,t))$, it will stay ``glued'' to the
particles at position $x-h$ and thus not jump.

Equation~\eqref{eq:markov-master} governs the evolution of the concentration
$\rho (x,t)$ on the discrete lattice. We perform a Taylor expansion in the
lattice spacing $h$ (see\ appendix) and consider the spatiotemporal scaling limit
\begin{equation} \label{eq:scaling}
h\downarrow 0,\quad \lambda _{0}\uparrow \infty , \quad h^{2}\lambda _{0}\rightarrow
D_{0}.
\end{equation}%
The particle concentration is then governed by the nonlinear advection-diffusion equation
\begin{equation}
\frac{\partial \rho }{\partial t}=\frac{\partial }{\partial x}\left[ D(\rho )%
\frac{\partial \rho }{\partial x}-\rho v(\rho )\right] 
\label{eq:markov-DE}
\end{equation}%
with diffusion coefficient $D(\rho )$ and drift coefficient $v(\rho )$ given
by
\begin{align*}
D(\rho )&=D_{0}\left[ a(\rho )q(\rho )+3a^{\prime }(\rho )q(\rho )\rho
-a(\rho )q^{\prime }(\rho )\rho \right] \\
v(\rho )&=-2D_{0}a(\rho )q(\rho )%
\frac{\partial S}{\partial x},
\end{align*}%
(note that $a^{\prime }(\cdot )$ and $q^{\prime }(\cdot )$ are plain
derivatives of the functions $a(\cdot )$ and $q(\cdot )$).
\cite{Anguige2011}, for instance,
studies volume filling and adhesion phenomena by setting
\begin{equation*}
q(\rho )=1-\rho ,\quad a(\rho )=1-\alpha \rho
\end{equation*}%
with adhesion parameter $\alpha >0$, resulting in
\begin{align*}
D(\rho )&=D_{0}\left[ 3\alpha \left( \rho -\frac{2}{3}\right) ^{2}+1-\frac{4}{%
3}\alpha \right] ,\\
v(\rho )&=-2D_{0}(1-\alpha \rho )(1-\rho )\frac{\partial S}{%
\partial x}.
\end{align*}

\section{Non-Markovian Transport}

\label{sec:non-markov}

This section is a short overview of fractional subdiffusion transport equations.
These equations have been successfully applied to
subdiffusive systems, whose main feature is a mean squared displacement of
sublinear growth $\sim t^{\mu }$ where $\mu \in (0,1)$. This is in stark
contrast to systems with Brownian noise, where mean squared displacement
grows linearly. The \textquotedblleft standard\textquotedblright\ fractional
diffusion equation governing the particle density $\rho (x,t)$ is
\begin{equation}
\frac{\partial ^{\mu }\rho }{\partial t^{\mu }}=D_{\mu }\frac{\partial
^{2}\rho }{\partial x^{2}},\quad \rho (x,0)=\rho _{0}(x),
\label{eq:standard-FDE}
\end{equation}%
where $D_{\mu }$ is a fractional diffusion constant with units length$^{2}$%
/time$^{\mu }$, and where the Caputo derivative of order $\mu $ is defined
via
\begin{equation}
\frac{\partial ^{\mu }}{\partial t^{\mu }}f(t):=\int_{0}^{t}f^{\prime }(t-s)%
\frac{s^{-\mu }}{\Gamma (1-\mu )}ds  \label{eq:frac-deriv}
\end{equation}%
with $\Gamma (\cdot )$ denoting the Gamma function. We note that in the
limit $\mu \uparrow 1$, the kernel $s^{-\mu }/\Gamma (1-\mu )$ converges to
the a Dirac delta, and the Caputo derivative is then the plain derivative of
order 1.
Similarly to the manner in which the standard diffusion equation with $\mu =1
$ is derived from a random walk, \eqref{eq:standard-FDE} is derived from a
Continuous Time Random Walk (CTRW) %
\citep{Metzler2000,limitCTRW,MeerschaertSikorskii}. Suppose that a particle
at the origin at time $t=0$ performs a random walk on a one-dimensional
lattice, and suppose that the waiting time $W$ between each jump is
distributed according to a power law with tail parameter $\mu \in (0,1)$:
\begin{equation} \label{eq:anomalous-exponent}
P(W>t)\sim \frac{(t/\tau _{0})^{-\mu }}{\Gamma (1-\mu )},\quad (t\rightarrow
\infty ).
\end{equation}%
Let $X(t)$ denote the random position of the particle at time $t$, and write
$p(x,t)$ for its probability density in space $x$ at time $t$.
Introducing the scaling parameters $h$ for space and $\tau_0$ for time, we
consider the rescaled position $hX(t/\tau _{0})$.
Its probability density is $p(x/h,t/\tau _{0})/h$.
Applying the scaling limit in which both $\tau_0$ and $h$ tend to $0$ and the ratio
$h^2/(2\tau_0^\mu)$ converges,
\begin{equation} \label{eq:scaling-anom}
h\downarrow 0, \quad \tau_{0}\downarrow 0, \quad \frac{h^{2}}{2\tau _{0}^{\mu }}%
\rightarrow D_{\mu },
\end{equation}%
this probability density also converges to a probability density $P(x,t)$ which solves the same equation as \eqref{eq:standard-FDE}. If one considers a collection of a large number of
particles with the above dynamics and assumes that the individual
trajectories do not interact, then $P(x,t)$ may be replaced by $\rho
(x,t)=CP(x,t)$, where $C$ denotes the total mass of particles. As %
\eqref{eq:standard-FDE} is linear, $\rho (x,t)$ is also a solution. It is
common to use this CTRW representation for Monte-Carlo simulations of
solutions of \eqref{eq:standard-FDE} %
\citep{MeerschaertSikorskii,timeLangevin}.

The above correspondence between CTRWs and fractional PDEs extends to CTRWs
with spatial variations, which can be applied to model chemotaxis problems %
\citep{HenryLanglands10chemo}. Suppose a collection of particles perform
independent CTRWs and respond to an external force $F(x)$ with a biased
jump probability, i.e.\ a probability $1/2+hF(x)$ to jump right and $%
1/2-hF(x)$ to jump left. Then a scaling limit as in \eqref{eq:scaling-anom}
yields a concentration $\rho (x,t)$ which is governed by the fractional
Fokker-Planck equation
\begin{equation}\label{eq:FFPEold}
\frac{\partial ^{\mu }\rho (x,t)}{\partial t^{\mu }}
=\frac{\partial^{2}}{\partial x^{2}}\left [D_{\mu }\rho (x,t)\right ]
-\frac{\partial }{\partial x}\left[ D_\mu F(x)\rho
(x,t)\right] ,
\quad \rho (x,0)=\rho _{0}(x),
\end{equation}%
Chemotaxis is usually modelled via a bias in the particle jumps which
depends on a chemotactic substance with concentration $S(x,t)$ according to $%
1/2\pm h\chi (\rho ,S)\partial S/\partial x$ %
\citep{Hillen2009,HenryLanglands10chemo}. Here the chemotactic sensitivity $%
\chi (\rho ,S)$ may be positive in the case of chemoattraction, or negative
in the case of chemorepulsion. This dynamics may be interpreted within the
fractional Fokker-Planck framework by letting the signal $F(\cdot )$ depend
on both space and time:
$$F(x,t)=\chi (\rho (x,t),S(x,t)) \frac{\partial S(x,t)}{\partial x}.$$
Importantly, however, Equation \eqref{eq:FFPEold} only
holds for external signals which do \emph{not} vary with time. The correct
generalization of the fractional Fokker-Planck equation for time-varying
signals $F=F(x,t)$ is
\begin{multline} \label{eq:FFPE}
\frac{\partial \rho (x,t)}{\partial t}
=\frac{\partial ^{2}}{\partial x^{2}} \left [ D_{\mu } \mathcal{D}_{t}^{1-\mu }\rho (x,t)\right ]
-\frac{\partial }{\partial x}\left[ D_\mu
F(x,t)\mathcal{D}_{t}^{1-\mu }\rho (x,t)\right],
\\
\rho (x,0)=\rho_{0}(x),
\end{multline}%
\citep{HLS10PRL} with order $1-\mu $ Riemann-Liouville fractional derivative
\begin{equation*}
\mathcal{D}_{t}^{1-\mu }f(t)=\frac{d}{dt}\int_{0}^{t}f(t-s)\frac{s^{\mu -1}}{%
\Gamma (\mu )}ds=\frac{\partial ^{1-\mu }f(t)}{\partial t^{1-\mu }}+f(0)%
\frac{t^{\mu -1}}{\Gamma (\mu )}.
\end{equation*}%

A further avenue of introducing spatial variations into fractional transport
is by letting the anomalous parameter vary in space, $\mu =\mu (x)$, see e.g.\ \cite{Chechkin2005a}. 
Equation \eqref{eq:anomalous-exponent} then becomes
\begin{equation*}
P(W>t)\sim \frac{(t/\tau _{0}(x))^{-\mu (x)}}{\Gamma (1-\mu (x))},\quad
(t\rightarrow \infty )
\end{equation*}%
\citep{Fedotov2012}.
For small $h$ and $\tau_0$,
the dynamics of the particle density $\rho (x,t)$ can be approximated by
\begin{multline} \label{eq:FFPE-nu-varies}
\frac{\partial \rho (x,t)}{\partial t}
=\frac{\partial^2 }{\partial x^2}\left[
D_{\mu }(x) \mathcal{D}_{t}^{1-\mu (x)}\rho (x,t)%
\right] 
-\frac{\partial }{\partial x}\left[ D_\mu(x) F(x,t)\mathcal{D}_{t}^{1-\mu
(x)}\rho (x,t)\right] ,\\ \rho (x,0)=\rho _{0}(x)
\end{multline}
where
$$D_{\mu }(x) = \frac{h^{2}}{2\tau_{0}^{\mu }(x)}.$$

Finally, we remark that many articles  study ``Caputo forms''
of \eqref{eq:FFPE} and \eqref{eq:FFPE-nu-varies},
which are obtained by simply substituting $F(x,t)$ for $F(x)$ and $\mu (x)$
for $\mu $ in \eqref{eq:FFPEold}. Such equations are not linked to
(Continuous Time) Random Walk models, and hence do not have any physical
interpretation. The authors deem it unlikely that a Caputo form of a transport
equation like \eqref{eq:FFPEold} can be derived from a chemotaxis model on
the lattice and strongly advocate the use of the Riemann-Liouville type equations %
\eqref{eq:FFPE} and \eqref{eq:FFPE-nu-varies}.

\section{Non-Markovian transport with nonlinear interaction}

This section contains the main results of our paper. 
We derive nonlinear fractional equations involving subdiffusion, volume filling and adhesion effects.
\label{sec:nonlin-nonmarkov} CTRWs whose waiting times between jumps are not
exponentially distributed do not satisfy the Markov property. However, they
are \emph{semi-Markov processes} \citep{Meerschaert2014} (or generalised renewal processes), meaning that the
Markov property applies at the random times at which a jump occurs. As a
consequence, at any time $t$ the law of the future trajectory of a particle
depends only on its position $x$ at time $t$ and the residence time $\tau$,
i.e.\ the time which has elapsed since the last jump. In mathematical terms,
the dynamics are Markovian on the state space $(x,\tau) \in \mathbb{R }%
\times [0,\infty)$.

As in \cite{Vlad2002a,Mendez2010}, we introduce the \emph{structured density}
$\xi(x,\tau,t) $ of particles at position $x$ at time $t$ whose residence
time equals $\tau$. Then the cell density $\rho(x,t)$ is recovered from the
structured density via simple integration:
\begin{align}  \label{u-xi}
\rho(x,t) = \int_0^\infty \xi(x,\tau,t) d\tau
\end{align}
As most other paper in the field, we assume the initial condition $%
\xi(x,\tau,0) = \rho_0(x) \delta(\tau) $ at time $t = 0 $. In this case, at
time $t $ no residence time can exceed the value $t $, and it suffices to
integrate over the domain $[0,t] $.

The structured density dynamics of a single particle are as follows: the
residence time $\tau $ increases linearly with time $t$ at the rate $1$,
until the particle escapes the site $x$. Upon escape, $\tau $ is reset to $0$%
. In this paper, we assume that the rate at which a particle escapes from a
site $x$ depends on two effects. Firstly, it depends on the (external)
environment at $x$ and neighbouring sites at time $t$, and secondly on its
(internal) residence time $\tau $ at $x$, which reflects a memory effect
typical for CTRWs. The external effect 
is comprised in the escape rate $\alpha (x,t)$, which may be a function
e.g.\ of the particle density $\rho $ at $x$ and neighbouring sites, and
thus may account for volume-exclusion and/or adhesion effects. Additionally,
it may be a function of the density $S(x,t)$ of a nearby chemically
signalling substance (see below). The internal effect, on the other hand, is
comprised in the escape rate $\gamma (x,\tau )$. For CTRWs, one typically
assumes a waiting time distribution with density $\psi (x,\tau )$ for the
times between jumps, with tail function $\Psi (x,\tau )=\int_{\tau }^{\infty
}\psi (x,\tau ^{\prime })d\tau ^{\prime }$. Then
\begin{equation}\label{eq:gamma-psi}
\gamma (x,\tau )=\psi (x,\tau )/\Psi (x,\tau )
\end{equation}
holds \citep{Fedotov2012}. A subdiffusive trapping effect at $x$ occurs when
$\gamma (x,\tau )$ is a decreasing function in $\tau $. For instance, a
Pareto distribution $\Psi (x,\tau )=(1+\tau /\tau _{0})^{-\beta }$ with
characteristic time scale $\tau _{0}$ and tail parameter $\mu \in (0,1)$
yields 
\begin{align} \label{eq:chemokinesis}
\gamma (x,\tau )=\mu /(\tau _{0}+\tau );
\end{align}
For the exponential
distribution $\Psi (x,\tau )=\exp (-t/\tau _{0})$, one has $\gamma (x,\tau
)=1/\tau _{0}$, independent of $\tau $, reflecting the typical lack of
memory.

For tractability, we assume that the internal and external effects are
independent. This means that the escape rates add up to a total escape rate $%
\alpha(x,t) + \gamma(x,\tau) $. The probability of an escape in the
infinitesimal time interval $(t,t+dt) $ is then $\alpha(x,t)dt +
\gamma(x,\tau) dt $. Now the dynamics of the structured density $%
\xi(x,\tau,t) $ can be seen to satisfy the equation
\begin{align}  \label{structured-dynamics}
\begin{split}
\frac{\partial }{\partial t} \xi(x,\tau,t) &= -\frac{\partial }{\partial
\tau } \xi(x,\tau,t) - [\alpha(x,t) + \gamma(x,\tau)] \xi(x,\tau,t), \quad t
> 0, \quad \tau > 0,
\end{split}%
\end{align}
which describes the linear increase of the residence time and the decay in
the structured particle density due to particle escapes.

Upon escape from $x$, the residence time $\tau $ is reset to $0$, and the
particle is placed back onto the lattice as follows: It jumps to the
neighbouring left resp.\ right lattice site with probability $L$ resp.\ $R$,
or it does not jump with probability $C$. We assume $L+R+C=1$.
Moreover, the probabilities $L$ and $R$ (and hence $C$) only depend on
external cues at $x$ at time $t$, and not on the internal residence time $%
\tau $ at the time of the jump: $L=L(x,t)$, $R=R(x,t)$. We incorporate
volume filling, adhesion and chemotactic drift into $L(x,t)$ and $R(x,t)$
via
\begin{equation*}
\begin{split}
L(x,t)& =q(\rho (x-h,t))a(\rho (x+h,t))
\left [\frac{1}{2} - \frac{S(x+h) - S(x-h)}{4}\right ]
\\
R(x,t)& =q(\rho (x+h,t))a(\rho (x-h,t))
\left [\frac{1}{2} + \frac{S(x+h) - S(x-h)}{4}\right ], \\
C(x,t)& =1-L(x,t)-R(x,t).
\end{split}%
\end{equation*}%
Here the functions $q(\rho )$, $a(\rho )$ and $S(x)$ play the same roles
(volume filling, adhesion and external signal) as described in \eqref{eq:Tpm-def}.
One can check that $L,R$ and $C$ are all probabilities, i.e.\ lie in the
interval $[0,1]$. 
Other choices for the impact of $S(x)$ on $L(x,t)$ and $R(x,t)$ are conceivable,
as in our derivation below we only assume that the bias equals
$
 \left [ 1 \pm h S'(x,t) + O(h^3) \right ] / 2. 
$
However the choice of terms $[1 + S(x\pm h) - S(x)]/2$ is not suitable, since we do not allow $L(x,t) + R(x,t) > 1$. 

The probabilities $L$, $R$ and $C$ thus define a dispersal
kernel
\begin{equation*}
w(x,t;z)=L(x,t)\delta (z+h)+R(x,t)\delta (z-h)+C(x,t)\delta (z),
\end{equation*}%
which is the probability distribution of a jump $z\in \{-h,+h,0\}$ given
that the jump occurs at time $t$ with base point $x$. If all particles with
density $\rho (x,t)$ are displaced according to $w(x,t;z)$, this results in
the density
\begin{equation*} 
\begin{split}
\mathcal{W}\rho (x,t)& :=\int_{z\in \mathbb{R}}\rho (x-z,t)w(x-z,t;z)dz \\
& =R(x-h,t)\rho (x-h,t)+L(x+h,t)\rho (x+h,t)+C(x,t)\rho (x,t).
\end{split}%
\end{equation*}%
The total escape flux from site $x$ at time $t$ is
\begin{align}\label{escape-flux}
\begin{split}
i(x,t)&:=\int_{0}^{\infty }[\alpha (x,t)+\gamma (x,\tau )]\xi (x,t,\tau
)d\tau 
\\
&=\alpha (x,t)\rho (x,t)+\int_{0}^{\infty }\gamma (x,\tau )\xi
(x,t,\tau )d\tau .
\end{split}
\end{align}%
Since all jumps are of nearest neighbor type, the quantity
\begin{equation*}
J\left( x+\frac{h}{2},t\right) :=R(x,t)i(x,t)h-L(x+h,t)i(x+h,t)h
\end{equation*}%
is readily interpreted as the net flux of particles from lattice point $x$
to lattice point $x+h$. Moreover, one confirms that
\begin{equation} \label{eq:discrete-flux}
h^{-1}\left[ J\left( x+\frac{h}{2},t\right) -J\left( x-\frac{h}{2},t\right) %
\right] =-\mathcal{W}i(x,t)+i(x,t).
\end{equation}%
By definition of $J(x,t)$ and conservation of mass, the above left-hand side
equals $-\partial \rho (x,t)/\partial t$. Assuming that the right-hand side
admits a valid Taylor expansion in the $x$-variable, we can write
\begin{equation} \label{eq:flux-transport}
\frac{\partial \rho (x,t)}{\partial t}=h^{2}\mathcal{A}i(x,t)+\mathcal{O}%
(h^{3}),
\end{equation}
where $\mathcal{A}$ is the transport operator
\begin{equation}
\begin{split}
\mathcal{A}i& =\frac{1}{2}\frac{\partial }{\partial x}\left[ a(\rho )q(\rho )%
\left[ \frac{\partial }{\partial x}i+i\left( \frac{3}{a(\rho )}\frac{%
\partial a(\rho )}{\partial x}-\frac{1}{q(\rho )}\frac{\partial q(\rho )}{%
\partial x}-2\frac{\partial S}{\partial x}\right) \right] \right] \\
& =\frac{1}{2}\frac{\partial }{\partial x}\left[ a(\rho )q(\rho )\left[
\frac{\partial }{\partial x}i+i\frac{\partial }{\partial x}\left( \log \frac{%
a(\rho )^{3}}{q(\rho )}-2S\right) \right] \right]
\end{split}
\label{transport}
\end{equation}%
acting on the $x$-variable only (see appendix). Recall the conservation of
mass equation $\partial \rho /\partial t+\partial J/\partial x=0$; The above
equation then allows for an interpretation of the flux $J(x,t)$ as the
decomposition into four components: (i) the local gradient of the escape
rate $i(x)$, (ii) adhesion effects due to $a(\rho )$, (iii) crowding
effects due to $q(\rho )$ and iv) the external signal $S(x,t)$. Equation %
\eqref{transport} will serve as the starting point for non-linear
transport equations, of both Markovian and time-fractional type, as we show
in the following two examples (it remains to express $i(x,t)$ in terms of $%
\rho (x,t)$).

\paragraph{Markovian nonlinear transport equations.}

If the residence time based escape rate $\gamma (x,\tau )$ vanishes and if $%
\alpha (x,t)=2\lambda _{0}$, \eqref{escape-flux} reads
\begin{equation}\label{eq:markovian-escape}
i(x,t)=2\lambda _{0}\rho (x,t).
\end{equation}
Now applying the diffusive scaling limit \eqref{eq:scaling}, we reproduce
the standard Markovian transport equation \eqref{eq:markov-DE}. In
particular, the flux equals
\begin{equation*}
J(x,t)=-D_{0}a(\rho )q(\rho )\left[ \frac{\partial \rho }{\partial x}+\rho
\frac{\partial }{\partial x}\left( \log \frac{a(\rho )^{3}}{q(\rho )}%
-2S\right) \right] ,
\end{equation*}%
where $D_{0}=h^{2}\lambda _{0}.$

\paragraph{Fractional nonlinear transport equations.}

Suppose now that $\alpha (x,t)$ vanishes and that the residence time based
escape rate is given by \eqref{eq:gamma-psi}, where
\begin{equation*}
\Psi (x,\tau )=E_{\mu }\left[ -(t/\tau _{0})^{\mu }\right]
\end{equation*}%
and $E_{\mu }$ denotes the Mittag-Leffler function (also see Table \ref%
{tab:waiting-times}). Then one has
\begin{equation}\label{eq:fractional-escape}
i(x,t)=\tau _{0}^{-\mu }\mathcal{D}_{t}^{1-\mu }\rho (x,t)
\end{equation}%
\citep[Eq.(30)]{Fedotov2012}. Applying the anomalous scaling limit %
\eqref{eq:scaling-anom}, the anomalous transport equation then equals
\begin{multline*}
\frac{\partial \rho (x,t)}{\partial t}=\frac{\partial }{\partial x}\left[
D_{\mu }a(\rho )q(\rho )\left[ \frac{\partial }{\partial x}\left( \mathcal{D}%
_{t}^{1-\mu }\rho \right) +\left( \mathcal{D}_{t}^{1-\mu }\rho \right) \frac{%
\partial }{\partial x}\left( \log \frac{a(\rho )^{3}}{q(\rho )}-2S\right) %
\right] \right], \\
\rho (x,0)=\rho _{0}(x),
\end{multline*}%
with general adhesion and volume filling effects $a(\rho (x,t))$ and $q(\rho
(x,t))$, external signal $S(x,t)$ and anomalous diffusion coefficient $D_{\mu
}=h^{2}/2\tau _{0}^{\mu }$. For instance, setting $q(\rho )\equiv 1$ (no
volume filling effect) and $a(\rho )=1-m\rho $ with adhesion parameter $m$
yields the fractional adhesion-diffusion equation
\begin{multline*}
\frac{\partial \rho }{\partial t}=\frac{\partial }{\partial x}\left[ D_{\mu
}(1-m\rho )\left[ \frac{\partial }{\partial x}\left( \mathcal{D}_{t}^{1-\mu
}\rho \right) +\left( \mathcal{D}_{t}^{1-\mu }\rho \right) \frac{\partial }{%
\partial x}\left( 3\log (1-m\rho )-2S\right) \right] \right] ,\\
\rho(x,0)=\rho _{0}(x).
\end{multline*}

\paragraph{Nonlinear transport equations with general memory effects.}

In the remainder of this section, we derive transport equations as
\eqref{eq:markovian-escape} and \eqref{eq:fractional-escape} in the case
where internal and external effects $\gamma(x,\tau)$ and $\alpha(x,t)$
coexist. This will provide models in which intermediate-time asymptotics are
subdiffusive, and long-time asymptotics are diffusive, see below. We begin
by solving the PDE \eqref{structured-dynamics}. For abbreviation, we
introduce the dimensionless functions
\begin{align*}
\Psi(x,t) &= \exp\left( -\int_0^t \gamma(x,s) ds \right), & \Phi(x,t) &=
\exp\left(-\int_0^t \alpha(x,s) ds \right)
\end{align*}
which take values in $(0,1]$. They may be interpreted as the probability
that in the time interval $[0,t] $ a particle has not escaped from $x$ due
to an internal (resp.\ external) effect. (Note that the above is consistent
with \eqref{eq:gamma-psi}.)
Assuming independence of the two effects, the probability that a particle
does not jump in the time interval $[0,t] $ is then $\Psi(x,t)\Phi(x,t) $.
We also note that
\begin{align*}
\psi(x,t) := -\frac{\partial }{\partial t} \Psi(x,t) = \gamma(x,t)
\Psi(x,t), \quad t > 0
\end{align*}
is a probability density. We write
\begin{align}  \label{eq:arrival}
j(x,t) := \xi(x,0,t)
\end{align}
for the flux of particles arriving at $x$.
A heuristic explanation for this interpretation is as follows:
The collection of particles at $x $ at time $t $ whose residence time lies
in the interval $(0,\varepsilon) $ have arrived there during the time
interval $(t-\varepsilon,t)$ and they have not escaped during this interval.
This balance equation reads
\begin{align*}
\int_0^\varepsilon \xi(x,\tau,t) d\tau = \int_{t-\varepsilon}^t j(x,s) \left[%
1- \int_s^t \alpha(x,r)dr - \int_s^t \gamma(x,r-s)dr + o(\varepsilon)\right]
ds.
\end{align*}
Now if we divide by $\varepsilon$ and let $\varepsilon \downarrow 0$ we
arrive at \eqref{eq:arrival}.

For simplicity, we assume that the initial structured density equals $\xi
(x,\tau ,0)=\rho _{0}(x)\delta (\tau )$, i.e.\ at time $0$ all particles
have residence time $0$ and their spatial distribution is $\rho _{0}(x)$.
Then we find via the method of characteristics (see appendix)
\begin{equation} \label{xi-cases}
\xi (x,\tau ,t)=\Psi (x,\tau )\dfrac{\Phi (x,t)}{\Phi (x,t-\tau )}j(x,t-\tau
)+\dfrac{\Psi (x,\tau )}{\Psi (x,\tau -t)}\Phi (x,t)\rho _{0}(x)\delta (\tau
-t).
\end{equation}%
Substituting this into \eqref{u-xi} and \eqref{escape-flux} gives the
equation pair
\begin{equation*}
\begin{split}
\rho (x,t)& =\int_{0}^{t}\Psi (x,\tau )\frac{\Phi (x,t)}{\Phi (x,t-\tau )}%
j(x,t-\tau )d\tau +\Psi (x,t)\Phi (x,t)\rho _{0}(x) \\
i(x,t)& =\alpha (x,t)\rho (x,t) \\
& +\int_{0}^{t}\psi (x,\tau )\frac{\Phi (x,t)}{\Phi (x,t-\tau )}j(x,t-\tau
)d\tau +\psi (x,t)\Phi (x,t)\rho _{0}(x).
\end{split}%
\end{equation*}%
We rewrite 
this in convenient shorthand notation:
\begin{align}
\frac{\rho }{\Phi }& =\Psi \ast _{t}\frac{j}{\Phi }+\Psi \rho _{0}
\label{eq:1} \\ \label{eq:2}
\frac{i}{\Phi }& =\frac{\alpha \rho }{\Phi }+\psi \ast _{t}\frac{j}{\Phi }%
+\psi \rho _{0}
\end{align}%
The symbol $\ast _{t}$ denotes a convolution in the time-variable $t$ (but
not in the space variable $x$). In order to derive an analytic form for the
escape rate $i(x,t)$, we introduce the function $m(x,t)$, defined via its
Laplace transform in $t$ as
\begin{equation} \label{eq:renewal-laplace}
\hat{m}(x,\lambda )=\int_{0}^{\infty }e^{-\lambda t}m(x,t)dt=\frac{\hat{\psi}%
(x,\lambda )}{1-\hat{\psi}(x,\lambda )}
\end{equation}
for any fixed $x$. This function is well-known in renewal theory as the
\emph{renewal measure density}\footnote{%
Although the renewal measure has an atom (singularity) at $0$, the renewal
measure density $t\mapsto m(x,t)$ does not have a delta function term at $0$.%
} associated with the probability density $t\mapsto \psi (x,t)$ %
\citep{feller1966introduction}. Its interpretation is that $%
\int_{a}^{b}m(x,t)dt$ equals the expected number of events (renewals) in the
time interval $(a,b]$, where the inter-arrival time of events is i.i.d.\
distributed with density $t\mapsto \psi (x,t)$. Two generic cases appear: if
$\psi (x,t)$ has finite first moment $\mu _{1}:=\int_{0}^{\infty }t\psi
(x,t)dt$, then for large times the rate of jumps evens out and approaches
the value $1/\mu _{1}$. In the case of a diverging first moment, i.e.\ $\mu
_{1}=\infty $, very long waiting times tend to occur, which means that for
very late times the rate of jumps decays to $0$. Four examples are collected
in Table \ref{tab:waiting-times}.
\begin{table}[t]
\centering
\begin{tabular}[h]{c|c|c|c}
				& $ \psi(t) $
				& $ \hat \psi(\lambda) $
				& $ m(t) $	\\
\hline\hline
Exponential  	& $ \tau_0^{-1} \exp(- t/\tau_0)  $
				& $ (1 + \tau_0 \lambda)^{-1}$
				& $ 1/\tau_0 $\\
\hline
Mittag-Leffler 	& $ -\frac{\del }{\del t} E_{\mu}\left[ - \left(\frac{t}{\tau_0} \right)^{\mu}  \right] $
				& $ \frac{1}{1+(\tau_0\lambda)^{\mu}} $
				& $ t^{\mu - 1} \tau_0^{-\mu} / \Gamma(\mu) $ \\
\hline
Gamma			& $ \frac{ t \exp(-t/\tau_0) }{\tau_0^{-2} }  $
				& $ (1 + \tau_0 \lambda)^{-2} $
				& $ \frac{1 - \exp(-2 t / \tau_0)}{2 \tau_0} $ \\
 \hline
 mixed Exp. 		& $ a_1 e^{-b_1 t} + a_2 e^{-b_2 t} $
 				& $ \frac{a_1}{c_1+\lambda}+\frac{a_2}{c_2+\lambda} $
 				& $ m_E(t) $
\end{tabular}
\caption{
\label{tab:waiting-times}
Waiting time distributions and their corresponding renewal measure densities.
The Mittag-Leffler density assumes $0 < \mu < 1$ and decays as $ t \to \infty $, according to a power-law $ \propto t^{-1-\mu} $.
The renewal measure density for the mixture of exponentials is
$m_{E}(t) = \frac{b_1 b_2}{b_1 a_2 + b_2 a_1} + \frac{(b_1 - b_2)^2 a_1 a_2}{b_1 a_2 + b_2 a_1} \times e^{-(b_1 a_2 + b_2 a_1)t}$
\citep[Problem III.5.2]{asmussen2003applied}.}
\end{table}
Now we can use Laplace transforms and the convolution formula
to show that \eqref{eq:1} is equivalent to
\begin{equation}\label{eq:RL}
\frac{\partial }{\partial t}\left( m\ast _{t}\frac{\rho }{\Phi }\right)
=\psi \ast _{t}\frac{j}{\Phi }+\psi \rho _{0}.
\end{equation}
Indeed, the Laplace transform of \eqref{eq:1} is 
\begin{align*}
\left (\dfrac{\rho}{\Phi}\right )^{\wedge}
= \hat \Psi \left (\dfrac{j}{\Phi}\right )^{\wedge}
+ \hat \Psi \rho_0
\end{align*}
whereas the Laplace transform of \eqref{eq:RL} is
\begin{align*}
\lambda  \hat m \left (\dfrac{\rho}{\Phi}\right )^\wedge 
= \hat \psi \left (\dfrac{j}{\Phi}\right)^\wedge
+ \hat \psi \rho_0,
\end{align*}
and the latter two equations are seen to be equivalent due to
\eqref{eq:renewal-laplace} and $1 - \lambda \hat \Psi = \hat \psi$.

Using \eqref{eq:2} then gives the result
\begin{equation*}
i=\alpha \rho +\Phi \,\frac{\partial }{\partial t}\left( \frac{\rho }{\Phi }%
\ast _{t}m\right)
\end{equation*}%
or, in detailed notation,
\begin{equation} \label{eq:rho2i-short}
i(x,t)=\alpha (x,t)\rho (x,t)+\Phi (x,t)\,\frac{\partial }{\partial t}%
\int_{0}^{t}\frac{\rho (x,s)}{\Phi (x,s)}m(t-s)ds,
\end{equation}%
where $\Phi (x,t)=\exp \left( -\int_{0}^{t}\alpha (x,s)ds\right) .$

The generalised Master equation takes the form
\begin{equation*}
\frac{\partial \rho (x,t)}{\partial t}=\frac{h^{2}}{2}\frac{\partial }{%
\partial x}\left[ a(\rho )q(\rho )\left[ \frac{\partial i}{\partial x}%
+i\left( \frac{3}{a(\rho )}\frac{\partial a(\rho )}{\partial x}-\frac{1}{%
q(\rho )}\frac{\partial q(\rho )}{\partial x}-2\frac{\partial S}{\partial x}%
\right) \right] \right] +\mathcal{O}(h^{3}),
\end{equation*}%
where as above $a(\rho (x,t))$ describes the adhesion effect,  $q(\rho(x,t))$ describes the volume filling effect and $S(x,t)$ is an external signal.  Let us consider a few examples illustrating the above equation. If we set $q(\rho)\equiv 1$ (no volume filling effect), $S=0$ and $a(\rho )=1-m\rho $ with adhesion parameter $m$, this yields the following master equation:
\begin{equation*}
\frac{\partial \rho (x,t)}{\partial t}=\frac{h^{2}}{2}\frac{\partial }{%
\partial x}\left[ (1-m\rho )\left[ \frac{\partial i}{\partial x}-i\left(
\frac{3m}{1-m\rho }\frac{\partial \rho }{\partial x}\right) \right] \right] +%
\mathcal{O}(h^{3})
\end{equation*}%
with
\begin{equation*}
i(x,t)=\alpha (x,t)\rho (x,t)+e^{-\int_{0}^{t}\alpha (x,s)ds}\,\frac{%
\partial }{\partial t}\int_{0}^{t}e^{\int_{0}^{s}\alpha (x,u)du}\rho
(x,s)m(t-s)ds.
\end{equation*}%
In the anomalous case, when the renewal measure density is
$$m(t)=t^{\mu
(x)-1}\tau _{0}^{-\mu (x)}/\Gamma (\mu (x)),$$ we can rewrite the last
expression for the total escape rate $i$ in terms of the fractional
derivative $\mathcal{D}_{t}^{1-\mu (x)}$:
\begin{equation} \label{eq:rho2i}
i(x,t)=\alpha (x,t)\rho (x,t)+\tau _{0}^{-\mu (x)}e^{-\int_{0}^{t}\alpha
(x,s)ds}\,\mathcal{D}_{t}^{1-\mu (x)}\left[ e^{\int_{0}^{t}\alpha
(x,u)du}\rho (x,t)\right] .
\end{equation}

Equation \eqref{eq:rho2i} is the sought generalisation of Equations %
\eqref{eq:markovian-escape} and \eqref{eq:fractional-escape}. The Markovian
situation \eqref{eq:markovian-escape} may be recovered by setting $\alpha
=2\lambda _{0}$ and $\gamma =0$, or equivalently by setting $\alpha =0$ and $%
\gamma =2\lambda _{0}$. The fractional situation \eqref{eq:fractional-escape}
results if $\alpha =0$ and if $\gamma (x,\tau )$ is as in %
\eqref{eq:gamma-psi}, where $\psi (x,\tau )$ at scale $\tau _{0}$ is
Mittag-Leffler (see Table~\ref{tab:waiting-times}).

Assume now the fractional situation as above, with the modification that $%
\alpha (x,t)$ be non-zero, finite and independent of the time scale
$\tau_{0}$. Taking the subdiffusive scaling limit \eqref{eq:scaling-anom} in
equation \eqref{eq:flux-transport} together with $\alpha \tau _{0}\ll 1$
results in the subdiffusive fractional evolution equation
\begin{equation} \label{eq:say-what???}
\frac{\partial \rho (x,t)}{\partial t}=2\mathcal{A}\left[ D_{\mu }(x)\Phi
(x,t)\,\mathcal{D}_{t}^{1-\mu (x)}\frac{\rho (x,t)}{\Phi (x,t)}\right]
\end{equation}%
where the transport operator $\mathcal{A}$ is defined in (\ref{transport}).
One can write this equation in the form
\begin{multline*}
\frac{\partial \rho (x,t)}{\partial t}
=\frac{\partial }{\partial x}\left\{
D_{\mu }(x)a(\rho )q(\rho )\frac{\partial }{\partial x}\left[ \Phi (x,t)\,%
\mathcal{D}_{t}^{1-\mu (x)}\frac{\rho (x,t)}{\Phi (x,t)}\right] \right\} \\
+\frac{\partial }{\partial x}\left\{ D_{\mu }(x)a(\rho )q(\rho )\left[ \Phi
(x,t)\,\mathcal{D}_{t}^{1-\mu }\frac{\rho (x,t)}{\Phi (x,t)}\right] \left(
\frac{3}{a(\rho )}\frac{\partial a(\rho )}{\partial x}-\frac{1}{q(\rho )}%
\frac{\partial q(\rho )}{\partial x}-2\frac{\partial S}{\partial x}\right)
\right\}
\end{multline*}
We note however that the PDE for the stationary solution has a much simpler form
\eqref{aggregation}.

\section{Aggregation phenomena in nonlinear subdiffusive systems}

\label{sec:profiles}

The purpose of this section is to analyse aggregation phenomena in subdiffusive systems,
which appear to be particularly intricate. \cite{Fedotov2012} have shown that a simple spatial variation in
the anomalous exponent $\mu $ (i.e.\ the power law exponent of the waiting
times) can cause the stationary profile to collapse, with all particles very
slowly aggregating at the one point where $\mu $ attains its minimum. In a
physical system, the particle count at any location will of course remain
bounded if particles have positive volumes, and such behaviour would be
deemed unphysical. With the developed theory, this \textquotedblleft volume
exclusion effect\textquotedblright\ can be taken into account for
subdiffusive aggregation.

%
%

\paragraph{Stationary structured density}

We look for necessary conditions for the structured density to yield a
stationary state. We assume that $t\mapsto \xi (x,\tau ,t)$ is constant for
every $(x,\tau )$; in this case the dependence on $t$ can be dropped, and we
write $\xi _{st}(x,\tau )$ for the stationary structured density. It follows
that the density $\rho $ and escape flux $i$ from \eqref{u-xi} and %
\eqref{escape-flux} are also stationary, and similarly we define $\rho
_{st}(x)$ and $i_{st}(x)$. Further assuming that $\alpha (x,t)=\alpha (x)$
does not depend on $t$, Equation \eqref{structured-dynamics} may now be
reinterpreted as
\begin{equation*}
\frac{\partial }{\partial \tau }\xi _{st}(x,\tau )=-[\alpha (x)+\gamma
(x,\tau )]\xi _{st}(x,\tau ),
\end{equation*}%
with solution
\begin{equation} \label{eq:stationary-structured-density}
\xi _{st}(x,\tau )=\xi _{st}(x,0)\exp (-\tau \lbrack \alpha (x)])\Psi
(x,\tau ).
\end{equation}
At equilibrium, the net flux $J(x,t)$ vanishes identically. Equations %
\eqref{eq:arrival}
and \eqref{eq:discrete-flux} then imply
$\xi _{st}(x,0)=j_{st}(x)=\mathcal{W}i_{st}(x)=i_{st}(x)$, and hence
\begin{equation*}
\xi _{st}(x,\tau )=i_{st}(x)\exp (-\tau \lbrack \alpha (x)])\Psi (x,\tau ).
\end{equation*}%
and integration over $\tau \in (0,\infty )$ yields
\begin{equation*}
\rho _{st}(x)=i_{st}(x)\hat{\Psi}(x,\alpha (x))
\end{equation*}%
where $\lambda \mapsto \hat{\Psi}(x,\lambda )$ denotes the Laplace transform
of $\tau \mapsto \Psi (x,\tau )$. We note that $\Psi (x,\alpha (x))$
equals the expected value of the random \textquotedblleft waiting
time\textquotedblright\ $T$ whose tail function $\mathbb{P}(T>\tau )$ equals
$\Psi (x,\tau )e^{-\tau \alpha (x)}$.
This is a kind of \emph{exponential tempering}
with tempering parameter $\alpha(x) $, a modification which ensures that all moments of a random waiting time are finite.
This is similar, but not identical to the tempering studied e.g.\ by \cite{Meerschaert2008Temper} and \cite{Stanislavsky2008}, where the factor $e^{-\tau\alpha(x)}$ is applied to $\psi(x,\tau)$ (and not $\Psi(x,\tau)$).
Finally, according to \eqref{eq:discrete-flux}, equilibrium holds if
\begin{equation*}
\mathcal{W}\left[ \frac{\rho _{st}(x)}{\hat{\Psi}(x,\alpha (x))}\right] -%
\frac{\rho _{st}(x)}{\hat{\Psi}(x,\alpha (x))}=0,
\end{equation*}%
which in the continuum limit becomes the aggregation equation
\begin{equation}
\mathcal{A}\left[ \frac{\rho _{st}(x)}{\hat{\Psi}(x,\alpha (x))}\right] =0,
\label{aggregation}
\end{equation}%
where the nonlinear transport operator $\mathcal{A}$ is defined in (\ref{transport}). This
equation is one of the main results of this paper. One can also write
\begin{equation*}
\frac{\partial }{\partial x}\left[ a(\rho _{st})q(\rho _{st})\left[ \frac{%
\partial }{\partial x}\left[ \frac{\rho _{st}(x)}{\hat{\Psi}(x,\alpha (x))}%
\right] +\left[ \frac{\rho _{st}(x)}{\hat{\Psi}(x,\alpha (x))}\right] \frac{%
\partial }{\partial x}\left( \log \frac{a(\rho _{st})^{3}}{q(\rho _{st})}%
+2S\right) \right] \right] =0.
\end{equation*}%
Apart from the transport operator $\mathcal{A}$ the stationary equation (\ref%
{aggregation}) involves a very important function $\hat{\Psi}(x,\alpha (x))$%
. Since
\begin{equation*}
\hat{\Psi}(x,\lambda )=\frac{1-\hat{\psi}(x,\lambda )}{\lambda },
\end{equation*}%
for the anomalous subdiffusive case with
\begin{equation*}
\hat{\psi}(x,\lambda )=\frac{1}{1+\left( \tau _{0}\lambda \right) ^{\mu (x)}}%
,
\end{equation*}%
we obtain
\begin{equation}
\frac{\rho _{st}(x)}{\hat{\Psi}(x,\alpha (x))}=\alpha (x)\rho _{st}(x)+\frac{%
\alpha (x)\rho _{st}(x)}{\left( \tau _{0}\alpha (x)\right) ^{\mu (x)}}.
\end{equation}%
When $\tau _{0}\alpha (x)$ is small, the second term becomes dominant and
determines the stationary profile $\rho _{st}(x)$ as a solution of the
equation
\begin{equation}
\mathcal{A}\left[ \frac{\alpha (x)\rho _{st}(x)}{\left( \tau _{0}\alpha
(x)\right) ^{\mu (x)}}\right] =0.
\end{equation}%
We should note that the equation (\ref{aggregation}) for the stationary
distribution $\rho _{st}(x)$ can not be obtained by simply equating the RHS of
the non-stationary master equation
\begin{equation*}
\frac{\partial \rho (x,t)}{\partial t}=h^{2}\mathcal{A}\left[ \alpha (x)\rho
(x,t)+e^{-\alpha (x)t}\,\frac{\partial }{\partial t}\int_{0}^{t}e^{\alpha
(x)s}\rho (x,s)m(t-s)ds\right]
\end{equation*}%
to $0$.
\begin{figure}
 \centering
 \includegraphics[width=.85\textwidth]{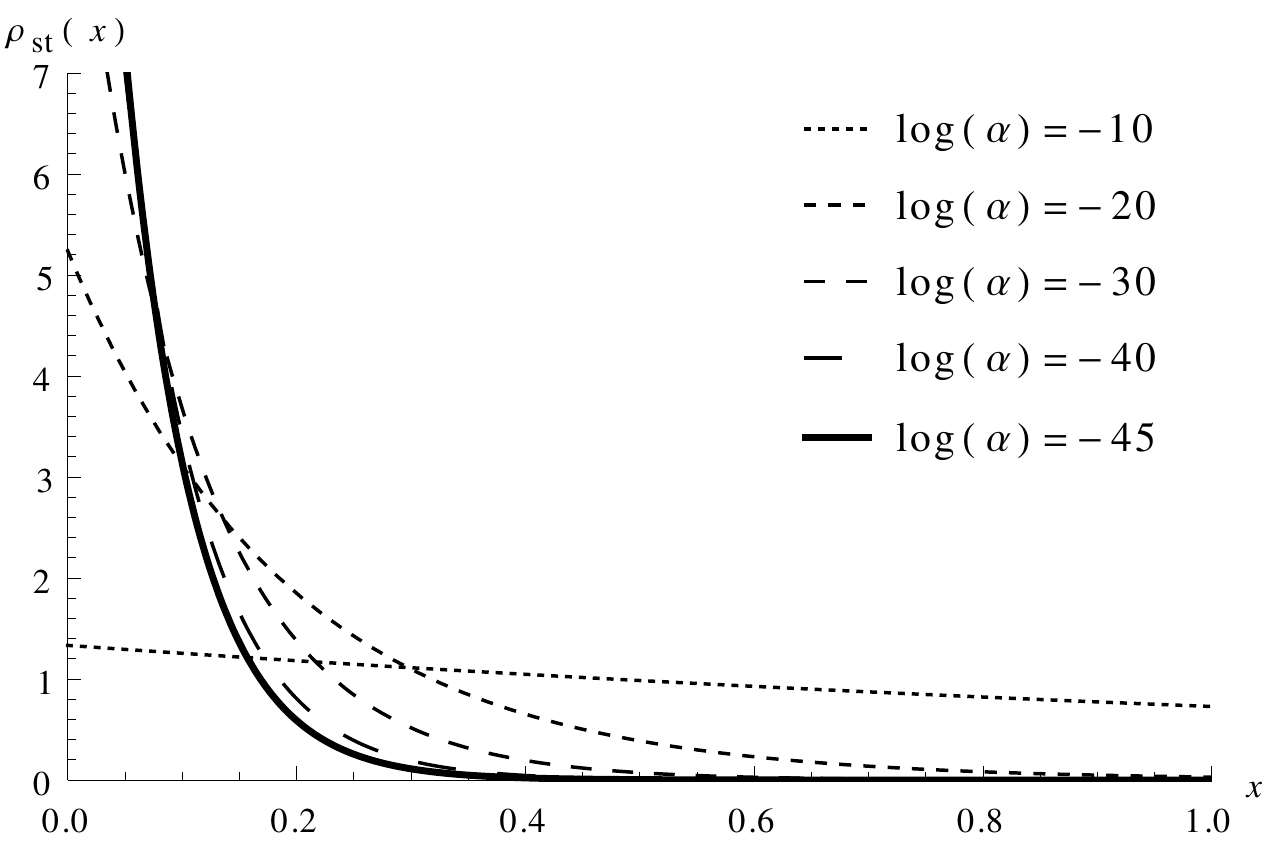}
 \caption{Aggregation of subdiffusive cells, 
 with $\mu(x) = 0.7 + 0.2x$, $S(x) = 2x$, 
 $\alpha(\rho) = 1$ and $q(\rho) = 1$.
 As $\alpha \downarrow 0$, the dynamics approach fractional dynamics.}
 \label{fig:no-vf}
\end{figure}
\begin{figure}
\centering
 \includegraphics[width=.85\textwidth]{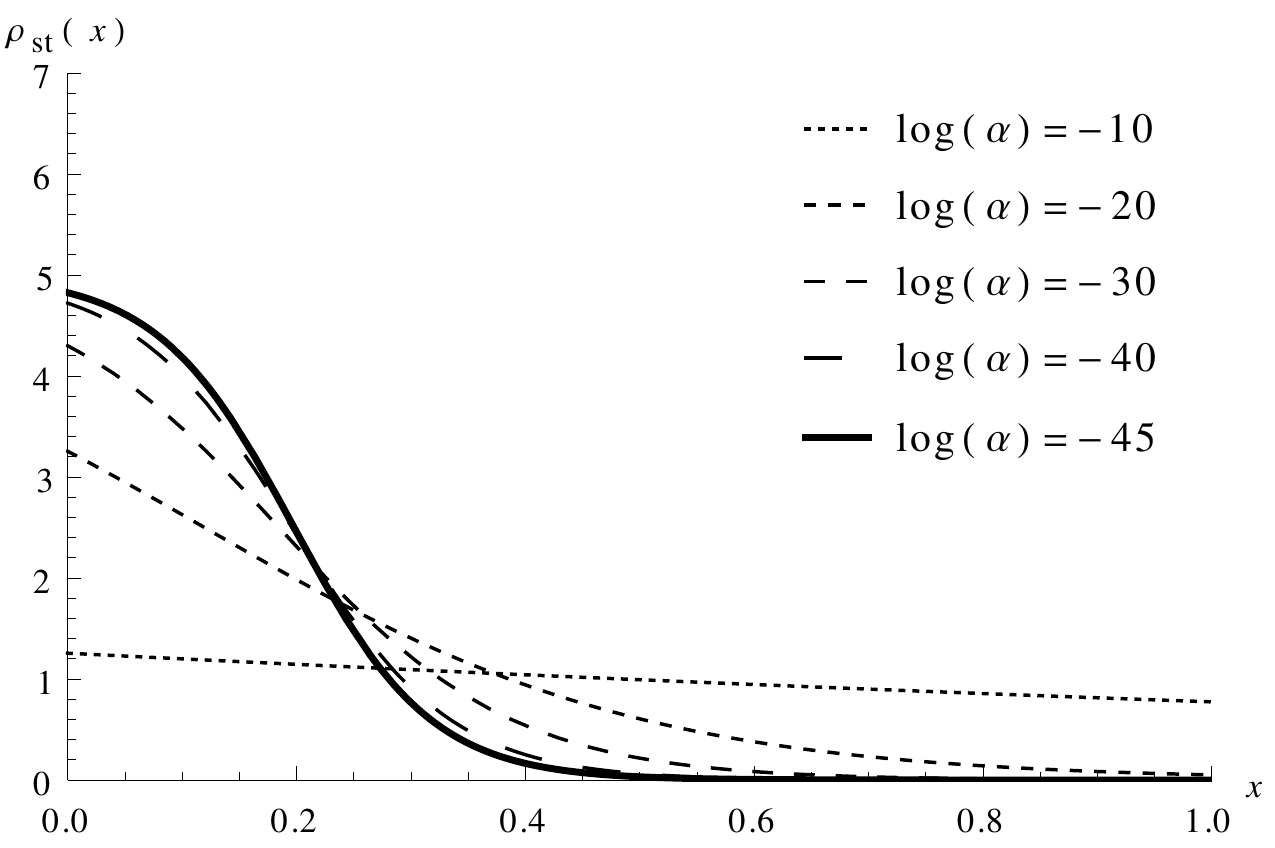}
 \caption{A system identical to Figure~\ref{fig:no-vf}, except with a nonlinear volume filling effect $q(\rho) = 1 - 0.2 \rho$. Anomalous aggregation is now visibly limited.}
 \label{fig:with-vf}
\end{figure}

Let us illustrate our general results by considering how nonlinear volume
filling effects interact with tempered anomalous aggregation. On the unit
interval $x\in \left[ 0,1\right] $, we find the stationary solutions to the equation (\ref{aggregation}) supposing that the anomalous
exponent is distributed as $\mu (x)=0.7+0.2x$, 
that the external signal equals $S(x) = 2x$ and that volume filling and adhesion effects are absent.
In the fractional case ($\alpha = 0$, see \eqref{eq:fractional-escape}), 
it is known that in the long time limit all particles accumulate at the minimum point of the anomalous exponent $\mu(x)$, independently of the initial configuration of the system \citep{Fedotov2012}. 
That is, the stationary density is singular, and in our case equals the delta function $\delta(x)$ at $0$. 
We illustrate this ``anomalous aggregation phenomenon'' again in Figure~\ref{fig:no-vf}, but in a different way: 
We first find the stationary solution $\rho_{st}(x)$ in the 
intermediate case \eqref{eq:rho2i} with $\alpha > 0$, which interpolates between fractional and Markovian dynamics.
We then let the parameter $\alpha$ tend to $0$, which means that $\rho_{st}(x)$ will approximate the stationary solution in the fractional case. 
(It should be noted that although the stationary distributions seemingly converge as $\alpha \downarrow 0$, 
the case $\alpha = 0$ is inherently different from the case $\alpha > 0$, because the stationary structured density $\xi_{st}(x,\tau)$ 
\eqref{eq:stationary-structured-density} only exists in the latter case.) 

We compute $\rho_{st}(x)$ in the intermediate case by solving \eqref{aggregation} with a nonlinear differential equation solver using Mathematica (see \ref{sec:app-stationary}). 
We do this for shrinking values of the parameter $\alpha$ and
indeed observe that $\rho_{st}(x)$ approaches, albeit slowly, the delta function; see Figure~\ref{fig:no-vf}. 
This occurs in spite of an external signal $S(x) = 2x$ which pushes the particles towards the right. 
We hence claim that anomalous aggregation is stronger than any (bounded) external signal $S(x)$. 

In Figure~\ref{fig:with-vf} we consider the same system, but with a volume filling effect $q(\rho) = 1 - 0.2 \rho$. 
Again, as $\alpha \downarrow 0$, the stationary solutions seem to converge. The limiting $\rho_{st}(x)$ however stays bounded below $5$ as $q(5) = 0$. 
We hence claim that nonlinear volume filling effects may effectively limit anomalous aggregation. 

\section{Conclusions}

The main challenge of this paper has been to implement nonlinear
effects such as volume filling and adhesion into fractional subdiffusive transport.
Starting with microscopic random walk models, we have derived \textit{non-Markovian and nonlinear} master equations for
the mean concentration of random walkers (cells, bacteria, etc.). We have taken into account
anomalous trapping, non-stationary tempering and nonlinear reactions together with nonlinear volume
filling and adhesion effects. We have shown that in the subdiffusive case these equations involve a nontrivial combination of the nonlinear terms together with fractional derivatives. The main point is that these equations can not be easily written phenomenologically. This is due to non-Markovian character of transport process involving anomalous trapping together with tempering. It turns out that in the long time limit these equations take a relatively simple form without fractional time derivatives  which allows to find the stationary solutions and thereby to study aggregation phenomena. We have shown that
nonlinear volume filling effects limit anomalous aggregation in subdiffusive transport systems with spatially nonuniform anomalous exponent.

\section*{Acknowledgements} 
The authors are grateful for support by the Engineering and Physical Sciences Research Council (EPSRC) through grant EP/J019526/1.
We thank Steven Falconer, Bruce Henry, Christopher Angstmann and Isaac Donnelly for helpful discussions. 

\section*{References}

\bibliographystyle{abbrvnat}
\bibliography{../../../Documents/Articles/library.bib}

\begin{thebibliography}{43}
\expandafter\ifx\csname natexlab\endcsname\relax\def\natexlab#1{#1}\fi
\providecommand{\url}[1]{\texttt{#1}}
\providecommand{\href}[2]{#2}
\providecommand{\path}[1]{#1}
\providecommand{\DOIprefix}{doi:}
\providecommand{\ArXivprefix}{arXiv:}
\providecommand{\URLprefix}{ }
\providecommand{\Pubmedprefix}{pmid:}
\providecommand{\doi}[1]{\href{http://dx.doi.org/#1}{\path{#1}}}
\providecommand{\Pubmed}[1]{\href{pmid:#1}{\path{#1}}}
\providecommand{\bibinfo}[2]{#2}
\ifx\xfnm\relax \def\xfnm[#1]{\unskip,\space#1}\fi
\bibitem[{Angstmann et~al.(2013)Angstmann, Donnelly \& Henry}]{Angstmann2013}
\bibinfo{author}{Angstmann, C.~N.}, \bibinfo{author}{Donnelly, I.~C.}, \&
  \bibinfo{author}{Henry, B.~I.} (\bibinfo{year}{2013}).
\newblock \bibinfo{title}{{Continuous Time Random Walks with Reactions Forcing
  and Trapping}}.
\newblock {\it \bibinfo{journal}{Math. Model. Nat. Phenom.}\/},  {\it
  \bibinfo{volume}{8}\/}, \bibinfo{pages}{17--27}.
  \DOIprefix\doi{10.1051/mmnp/20138202}.
\bibitem[{Anguige(2011)}]{Anguige2011}
\bibinfo{author}{Anguige, K.} (\bibinfo{year}{2011}).
\newblock \bibinfo{title}{{A one-dimensional model for the interaction between
  cell-to-cell adhesion and chemotactic signalling}}.
\newblock {\it \bibinfo{journal}{Eur. J. Appl. Math.}\/},  {\it
  \bibinfo{volume}{22}\/}, \bibinfo{pages}{291--316}.
  \DOIprefix\doi{10.1017/S0956792511000040}.
\bibitem[{Anguige \& Schmeiser(2009)}]{Anguige2009}
\bibinfo{author}{Anguige, K.}, \& \bibinfo{author}{Schmeiser, C.}
  (\bibinfo{year}{2009}).
\newblock \bibinfo{title}{{A one-dimensional model of cell diffusion and
  aggregation, incorporating volume filling and cell-to-cell adhesion.}}
\newblock {\it \bibinfo{journal}{J. Math. Biol.}\/},  {\it
  \bibinfo{volume}{58}\/}, \bibinfo{pages}{395--427}.
  \DOIprefix\doi{10.1007/s00285-008-0197-8}.
\bibitem[{Armstrong et~al.(2006)Armstrong, Painter \& Sherratt}]{Armstrong2006}
\bibinfo{author}{Armstrong, N.~J.}, \bibinfo{author}{Painter, K.~J.}, \&
  \bibinfo{author}{Sherratt, J.~A.} (\bibinfo{year}{2006}).
\newblock \bibinfo{title}{{A continuum approach to modelling cell-cell
  adhesion.}}
\newblock {\it \bibinfo{journal}{J. Theor. Biol.}\/},  {\it
  \bibinfo{volume}{243}\/}, \bibinfo{pages}{98--113}.
  \DOIprefix\doi{10.1016/j.jtbi.2006.05.030}.
\bibitem[{Asmussen(2003)}]{asmussen2003applied}
\bibinfo{author}{Asmussen, S.} (\bibinfo{year}{2003}).
\newblock {\it \bibinfo{title}{{Applied probability and queues}}\/}
  volume~\bibinfo{volume}{2}.
\newblock \bibinfo{publisher}{Springer New York}.
\bibitem[{Banks \& Fradin(2005)}]{Banks2005}
\bibinfo{author}{Banks, D.~S.}, \& \bibinfo{author}{Fradin, C.}
  (\bibinfo{year}{2005}).
\newblock \bibinfo{title}{{Anomalous diffusion of proteins due to molecular
  crowding.}}
\newblock {\it \bibinfo{journal}{Biophys. J.}\/},  {\it
  \bibinfo{volume}{89}\/}, \bibinfo{pages}{2960--71}.
  \DOIprefix\doi{10.1529/biophysj.104.051078}.
\bibitem[{Barkai et~al.(2000)Barkai, Metzler \& Klafter}]{BMK00}
\bibinfo{author}{Barkai, E.}, \bibinfo{author}{Metzler, R.}, \&
  \bibinfo{author}{Klafter, J.} (\bibinfo{year}{2000}).
\newblock \bibinfo{title}{{From continuous time random walks to the fractional
  Fokker-Planck equation}}.
\newblock {\it \bibinfo{journal}{Phys. Rev. E}\/},  {\it
  \bibinfo{volume}{61}\/}, \bibinfo{pages}{132--138}.
  \DOIprefix\doi{10.1103/PhysRevE.61.132}.
\bibitem[{Chechkin et~al.(2005)Chechkin, Gorenflo \& Sokolov}]{Chechkin2005a}
\bibinfo{author}{Chechkin, A.~V.}, \bibinfo{author}{Gorenflo, R.}, \&
  \bibinfo{author}{Sokolov, I.~M.} (\bibinfo{year}{2005}).
\newblock \bibinfo{title}{{Fractional diffusion in inhomogeneous media}}.
\newblock {\it \bibinfo{journal}{J. Phys. A. Math. Gen.}\/},  {\it
  \bibinfo{volume}{38}\/}, \bibinfo{pages}{L679--L684}.
  \DOIprefix\doi{10.1088/0305-4470/38/42/L03}.
\bibitem[{Dieterich et~al.(2008)Dieterich, Klages, Preuss \&
  Schwab}]{Dieterich2008}
\bibinfo{author}{Dieterich, P.}, \bibinfo{author}{Klages, R.},
  \bibinfo{author}{Preuss, R.}, \& \bibinfo{author}{Schwab, A.}
  (\bibinfo{year}{2008}).
\newblock \bibinfo{title}{{Anomalous dynamics of cell migration}}.
\newblock {\it \bibinfo{journal}{Proc. Natl. Acad. Sci. U. S. A.}\/},  {\it
  \bibinfo{volume}{105}\/}, \bibinfo{pages}{459--463}.
  \DOIprefix\doi{10.1073/pnas.0707603105}.
\bibitem[{Erban \& Othmer(2005)}]{erban2005signal}
\bibinfo{author}{Erban, R.}, \& \bibinfo{author}{Othmer, H.~G.}
  (\bibinfo{year}{2005}).
\newblock \bibinfo{title}{{From Signal Transduction to Spatial Pattern
  Formation in E. coli: A Paradigm for Multiscale Modeling in Biology}}.
\newblock {\it \bibinfo{journal}{Multiscale Model. Simul.}\/},  {\it
  \bibinfo{volume}{3}\/}, \bibinfo{pages}{362--394}.
  \DOIprefix\doi{10.1137/040603565}.
\bibitem[{Fedotov(2013)}]{Fedotov2013b}
\bibinfo{author}{Fedotov, S.} (\bibinfo{year}{2013}).
\newblock \bibinfo{title}{{Nonlinear subdiffusive fractional equations and the
  aggregation phenomenon}}.
\newblock {\it \bibinfo{journal}{Phys. Rev. E}\/},  {\it
  \bibinfo{volume}{88}\/}, \bibinfo{pages}{32104}.
  \DOIprefix\doi{10.1103/PhysRevE.88.032104}.
\bibitem[{Fedotov \& Falconer(2012)}]{Fedotov2012}
\bibinfo{author}{Fedotov, S.}, \& \bibinfo{author}{Falconer, S.}
  (\bibinfo{year}{2012}).
\newblock \bibinfo{title}{{Subdiffusive master equation with space-dependent
  anomalous exponent and structural instability}}.
\newblock {\it \bibinfo{journal}{Phys. Rev. E}\/},  {\it
  \bibinfo{volume}{85}\/}, \bibinfo{pages}{031132}.
  \DOIprefix\doi{10.1103/PhysRevE.85.031132}.
\bibitem[{Fedotov et~al.(2013)Fedotov, Ivanov \& Zubarev}]{Fedotov2013a}
\bibinfo{author}{Fedotov, S.}, \bibinfo{author}{Ivanov, A.~O.}, \&
  \bibinfo{author}{Zubarev, A.~Y.} (\bibinfo{year}{2013}).
\newblock \bibinfo{title}{{Non-homogeneous Random Walks, Subdiffusive Migration
  of Cells and Anomalous Chemotaxis}}.
\newblock {\it \bibinfo{journal}{Math. Model. Nat. Phenom.}\/},  {\it
  \bibinfo{volume}{8}\/}, \bibinfo{pages}{28--43}.
  \DOIprefix\doi{10.1051/mmnp/20138203}.
\bibitem[{Feller(1966)}]{feller1966introduction}
\bibinfo{author}{Feller, W.} (\bibinfo{year}{1966}).
\newblock \bibinfo{title}{{An introduction to probability theory, Vol. II}}.
\bibitem[{Fenchel \& Blackburn(1999)}]{fenchel1999motile}
\bibinfo{author}{Fenchel, T.}, \& \bibinfo{author}{Blackburn, N.}
  (\bibinfo{year}{1999}).
\newblock \bibinfo{title}{{Motile chemosensory behaviour of phagotrophic
  protists: mechanisms for and efficiency in congregating at food patches.}}
\newblock {\it \bibinfo{journal}{Protist}\/},  {\it \bibinfo{volume}{150}\/},
  \bibinfo{pages}{325--36}. \DOIprefix\doi{10.1016/S1434-4610(99)70033-7}.
\bibitem[{Fernando et~al.(2010)Fernando, Landman \& Simpson}]{Fernando2010}
\bibinfo{author}{Fernando, A.~E.}, \bibinfo{author}{Landman, K.~A.}, \&
  \bibinfo{author}{Simpson, M.~J.} (\bibinfo{year}{2010}).
\newblock \bibinfo{title}{{Nonlinear diffusion and exclusion processes with
  contact interactions}}.
\newblock {\it \bibinfo{journal}{Phys. Rev. E}\/},  {\it
  \bibinfo{volume}{81}\/}, \bibinfo{pages}{011903}.
  \DOIprefix\doi{10.1103/PhysRevE.81.011903}.
\bibitem[{Golding \& Cox(2006)}]{Golding2006}
\bibinfo{author}{Golding, I.}, \& \bibinfo{author}{Cox, E.}
  (\bibinfo{year}{2006}).
\newblock \bibinfo{title}{{Physical Nature of Bacterial Cytoplasm}}.
\newblock {\it \bibinfo{journal}{Phys. Rev. Lett.}\/},  {\it
  \bibinfo{volume}{96}\/}, \bibinfo{pages}{098102}.
  \DOIprefix\doi{10.1103/PhysRevLett.96.098102}.
\bibitem[{Henry et~al.(2006)Henry, Langlands \& Wearne}]{Henry2006}
\bibinfo{author}{Henry, B.}, \bibinfo{author}{Langlands, T.}, \&
  \bibinfo{author}{Wearne, S.} (\bibinfo{year}{2006}).
\newblock \bibinfo{title}{{Anomalous diffusion with linear reaction dynamics:
  From continuous time random walks to fractional reaction-diffusion
  equations}}.
\newblock {\it \bibinfo{journal}{Phys. Rev. E}\/},  {\it
  \bibinfo{volume}{74}\/}, \bibinfo{pages}{031116}.
  \DOIprefix\doi{10.1103/PhysRevE.74.031116}.
\bibitem[{Henry et~al.(2010)Henry, Langlands \& Straka}]{HLS10PRL}
\bibinfo{author}{Henry, B.~I.}, \bibinfo{author}{Langlands, T. A.~M.}, \&
  \bibinfo{author}{Straka, P.} (\bibinfo{year}{2010}).
\newblock \bibinfo{title}{{Fractional Fokker-Planck Equations for Subdiffusion
  with Space- and Time-Dependent Forces}}.
\newblock {\it \bibinfo{journal}{Phys. Rev. Lett.}\/},  {\it
  \bibinfo{volume}{105}\/}, \bibinfo{pages}{170602}.
  \DOIprefix\doi{10.1103/PhysRevLett.105.170602}.
\bibitem[{Hillen \& Painter(2009)}]{Hillen2009}
\bibinfo{author}{Hillen, T.}, \& \bibinfo{author}{Painter, K.~J.}
  (\bibinfo{year}{2009}).
\newblock \bibinfo{title}{{A user's guide to PDE models for chemotaxis.}}
\newblock {\it \bibinfo{journal}{J. Math. Biol.}\/},  {\it
  \bibinfo{volume}{58}\/}, \bibinfo{pages}{183--217}.
  \DOIprefix\doi{10.1007/s00285-008-0201-3}.
\bibitem[{Johnston et~al.(2012)Johnston, Simpson \& Baker}]{Johnston2012}
\bibinfo{author}{Johnston, S.~T.}, \bibinfo{author}{Simpson, M.~J.}, \&
  \bibinfo{author}{Baker, R.~E.} (\bibinfo{year}{2012}).
\newblock \bibinfo{title}{{Mean-field descriptions of collective migration with
  strong adhesion}}.
\newblock {\it \bibinfo{journal}{Phys. Rev. E}\/},  {\it
  \bibinfo{volume}{85}\/}, \bibinfo{pages}{051922}.
  \DOIprefix\doi{10.1103/PhysRevE.85.051922}.
\bibitem[{Langlands \& Henry(2010)}]{HenryLanglands10chemo}
\bibinfo{author}{Langlands, T. A.~M.}, \& \bibinfo{author}{Henry, B.}
  (\bibinfo{year}{2010}).
\newblock \bibinfo{title}{{Fractional chemotaxis diffusion equations}}.
\newblock {\it \bibinfo{journal}{Phys. Rev. E}\/},  {\it
  \bibinfo{volume}{81}\/}, \bibinfo{pages}{051102}.
  \DOIprefix\doi{10.1103/PhysRevE.81.051102}.
\bibitem[{Meerschaert \& Scheffler(2004)}]{limitCTRW}
\bibinfo{author}{Meerschaert, M.~M.}, \& \bibinfo{author}{Scheffler, H.-P.}
  (\bibinfo{year}{2004}).
\newblock \bibinfo{title}{{Limit theorems for continuous-time random walks with
  infinite mean waiting times}}.
\newblock {\it \bibinfo{journal}{J. Appl. Probab.}\/},  {\it
  \bibinfo{volume}{41}\/}, \bibinfo{pages}{623--638}.
  \DOIprefix\doi{10.1239/jap/1091543414}.
\bibitem[{Meerschaert \& Sikorskii(2011)}]{MeerschaertSikorskii}
\bibinfo{author}{Meerschaert, M.~M.}, \& \bibinfo{author}{Sikorskii, A.}
  (\bibinfo{year}{2011}).
\newblock {\it \bibinfo{title}{{Stochastic models for fractional calculus}}\/}.
\newblock \bibinfo{publisher}{De Gruyter}.
\bibitem[{Meerschaert \& Straka(2014)}]{Meerschaert2014}
\bibinfo{author}{Meerschaert, M.~M.}, \& \bibinfo{author}{Straka, P.}
  (\bibinfo{year}{2014}).
\newblock \bibinfo{title}{{Semi-Markov approach to continuous time random walk
  limit processes}}.
\newblock {\it \bibinfo{journal}{Ann. Probab.}\/},  {\it
  \bibinfo{volume}{42}\/}, \bibinfo{pages}{1699--1723}.
  \DOIprefix\doi{10.1214/13-AOP905}.
\bibitem[{Meerschaert et~al.(2008)Meerschaert, Zhang \&
  Baeumer}]{Meerschaert2008Temper}
\bibinfo{author}{Meerschaert, M.~M.}, \bibinfo{author}{Zhang, Y.}, \&
  \bibinfo{author}{Baeumer, B.} (\bibinfo{year}{2008}).
\newblock \bibinfo{title}{{Tempered anomalous diffusion in heterogeneous
  systems}}.
\newblock {\it \bibinfo{journal}{Geophys. Res. Lett.}\/},  {\it
  \bibinfo{volume}{35}\/}, \bibinfo{pages}{L17403}.
  \DOIprefix\doi{10.1029/2008GL034899}.
\bibitem[{M\'{e}ndez et~al.(2012)M\'{e}ndez, Campos, Pagonabarraga \&
  Fedotov}]{Mendez2012}
\bibinfo{author}{M\'{e}ndez, V.}, \bibinfo{author}{Campos, D.},
  \bibinfo{author}{Pagonabarraga, I.}, \& \bibinfo{author}{Fedotov, S.}
  (\bibinfo{year}{2012}).
\newblock \bibinfo{title}{{Density-dependent dispersal and population
  aggregation patterns}}.
\newblock {\it \bibinfo{journal}{J. Theor. Biol.}\/},  {\it
  \bibinfo{volume}{309}\/}, \bibinfo{pages}{113--20}.
  \DOIprefix\doi{10.1016/j.jtbi.2012.06.015}.
\bibitem[{Mendez et~al.(2010)Mendez, Fedotov \& Horsthemke}]{Mendez2010}
\bibinfo{author}{Mendez, V.}, \bibinfo{author}{Fedotov, S.}, \&
  \bibinfo{author}{Horsthemke, W.} (\bibinfo{year}{2010}).
\newblock {\it \bibinfo{title}{{Reaction-Transport Systems: Mesoscopic
  Foundations, Fronts, and Spatial Instabilities}}\/}.
\newblock (\bibinfo{edition}{1st} ed.).
\newblock \bibinfo{publisher}{Springer}.
\bibitem[{Metzler \& Klafter(2000)}]{Metzler2000}
\bibinfo{author}{Metzler, R.}, \& \bibinfo{author}{Klafter, J.}
  (\bibinfo{year}{2000}).
\newblock \bibinfo{title}{{The random walk's guide to anomalous diffusion: a
  fractional dynamics approach}}.
\newblock {\it \bibinfo{journal}{Phys. Rep.}\/},  {\it
  \bibinfo{volume}{339}\/}, \bibinfo{pages}{1--77}.
  \DOIprefix\doi{10.1016/S0370-1573(00)00070-3}.
\bibitem[{Mierke et~al.(2011)Mierke, Frey, Fellner, Herrmann \&
  Fabry}]{mierke2011integrin}
\bibinfo{author}{Mierke, C.~T.}, \bibinfo{author}{Frey, B.},
  \bibinfo{author}{Fellner, M.}, \bibinfo{author}{Herrmann, M.}, \&
  \bibinfo{author}{Fabry, B.} (\bibinfo{year}{2011}).
\newblock \bibinfo{title}{{Integrin $\alpha$5$\beta$1 facilitates cancer cell
  invasion through enhanced contractile forces}}.
\newblock {\it \bibinfo{journal}{J. Cell Sci.}\/},  {\it
  \bibinfo{volume}{124}\/}, \bibinfo{pages}{369--83}.
  \DOIprefix\doi{10.1242/jcs.071985}.
\bibitem[{Murray(2007)}]{Murray2002}
\bibinfo{author}{Murray, J.} (\bibinfo{year}{2007}).
\newblock {\it \bibinfo{title}{{Mathematical Biology: I. An Introduction}}\/}.
\newblock Interdisciplinary applied mathematics (\bibinfo{edition}{3rd} ed.).
\newblock \bibinfo{publisher}{Springer}.
\bibitem[{Oelschl\"{a}ger(1989)}]{Oelschlager1989}
\bibinfo{author}{Oelschl\"{a}ger, K.} (\bibinfo{year}{1989}).
\newblock \bibinfo{title}{{On the derivation of reaction-diffusion equations as
  limit dynamics of systems of moderately interacting stochastic processes}}.
\newblock {\it \bibinfo{journal}{Probab. Theory Relat. Fields}\/},  {\it
  \bibinfo{volume}{82}\/}, \bibinfo{pages}{565--586}.
  \DOIprefix\doi{10.1007/BF00341284}.
\bibitem[{Othmer \& Hillen(2002)}]{Othmer2002a}
\bibinfo{author}{Othmer, H.~G.}, \& \bibinfo{author}{Hillen, T.}
  (\bibinfo{year}{2002}).
\newblock \bibinfo{title}{{The Diffusion Limit of Transport Equations II:
  Chemotaxis Equations}}.
\newblock {\it \bibinfo{journal}{SIAM J. Appl. Math.}\/},  {\it
  \bibinfo{volume}{62}\/}, \bibinfo{pages}{1222--1250}.
  \DOIprefix\doi{10.1137/S0036139900382772}.
\bibitem[{Painter \& Hillen(2002)}]{Painter2002}
\bibinfo{author}{Painter, K.~J.}, \& \bibinfo{author}{Hillen, T.}
  (\bibinfo{year}{2002}).
\newblock \bibinfo{title}{{Volume-filling and quorum-sensing in models for
  chemosensitive movement}}.
\newblock {\it \bibinfo{journal}{Can. Appl. Math. Quart}\/},  {\it
  \bibinfo{volume}{10}\/}, \bibinfo{pages}{1--32}.
\bibitem[{Simpson \& Baker(2011)}]{Simpson2011}
\bibinfo{author}{Simpson, M.~J.}, \& \bibinfo{author}{Baker, R.~E.}
  (\bibinfo{year}{2011}).
\newblock \bibinfo{title}{{Corrected mean-field models for spatially dependent
  advection-diffusion-reaction phenomena}}.
\newblock {\it \bibinfo{journal}{Phys. Rev. E}\/},  {\it
  \bibinfo{volume}{83}\/}, \bibinfo{pages}{051922}.
  \DOIprefix\doi{10.1103/PhysRevE.83.051922}.
\bibitem[{Stanislavsky et~al.(2008)Stanislavsky, Weron \&
  Weron}]{Stanislavsky2008}
\bibinfo{author}{Stanislavsky, A.}, \bibinfo{author}{Weron, K.}, \&
  \bibinfo{author}{Weron, A.} (\bibinfo{year}{2008}).
\newblock \bibinfo{title}{{Diffusion and relaxation controlled by tempered
  $\alpha$-stable processes}}.
\newblock {\it \bibinfo{journal}{Phys. Rev. E}\/},  {\it
  \bibinfo{volume}{78}\/}, \bibinfo{pages}{051106}.
  \DOIprefix\doi{10.1103/PhysRevE.78.051106}.
\bibitem[{Stevens(2000)}]{Stevens00}
\bibinfo{author}{Stevens, A.} (\bibinfo{year}{2000}).
\newblock \bibinfo{title}{{The Derivation of Chemotaxis Equations as Limit
  Dynamics of Moderately Interacting Stochastic Many-Particle Systems}}.
\newblock {\it \bibinfo{journal}{SIAM J. Appl. Math.}\/},  {\it
  \bibinfo{volume}{61}\/}, \bibinfo{pages}{183--212}.
  \DOIprefix\doi{10.1137/S0036139998342065}.
\bibitem[{Stevens \& Othmer(1997)}]{Stevens1997a}
\bibinfo{author}{Stevens, A.}, \& \bibinfo{author}{Othmer, H.~G.}
  (\bibinfo{year}{1997}).
\newblock \bibinfo{title}{{Aggregation, Blowup, and Collapse: The ABC's of
  Taxis in Reinforced Random Walks}}.
\newblock {\it \bibinfo{journal}{SIAM J. Appl. Math.}\/},  {\it
  \bibinfo{volume}{57}\/}, \bibinfo{pages}{1044--1081}.
  \DOIprefix\doi{10.1137/S0036139995288976}.
\bibitem[{Toli\'{c}-N\o{}rrelykke et~al.(2004)Toli\'{c}-N\o{}rrelykke,
  Munteanu, Thon, Oddershede \& Berg-S\o~rensen}]{TMT04}
\bibinfo{author}{Toli\'{c}-N\o{}rrelykke, I.}, \bibinfo{author}{Munteanu,
  E.-L.}, \bibinfo{author}{Thon, G.}, \bibinfo{author}{Oddershede, L.}, \&
  \bibinfo{author}{Berg-S\o~rensen, K.} (\bibinfo{year}{2004}).
\newblock \bibinfo{title}{{Anomalous Diffusion in Living Yeast Cells}}.
\newblock {\it \bibinfo{journal}{Phys. Rev. Lett.}\/},  {\it
  \bibinfo{volume}{93}\/}, \bibinfo{pages}{078102}.
  \DOIprefix\doi{10.1103/PhysRevLett.93.078102}.
\bibitem[{Vlad \& Ross(2002)}]{Vlad2002a}
\bibinfo{author}{Vlad, M.}, \& \bibinfo{author}{Ross, J.}
  (\bibinfo{year}{2002}).
\newblock \bibinfo{title}{{Systematic derivation of reaction-diffusion
  equations with distributed delays and relations to fractional
  reaction-diffusion equations and hyperbolic transport equations: Application
  to the theory of Neolithic transition}}.
\newblock {\it \bibinfo{journal}{Phys. Rev. E}\/},  {\it
  \bibinfo{volume}{66}\/}, \bibinfo{pages}{061908}.
  \DOIprefix\doi{10.1103/PhysRevE.66.061908}.
\bibitem[{Wadhams \& Armitage(2004)}]{wadhams2004making}
\bibinfo{author}{Wadhams, G.~H.}, \& \bibinfo{author}{Armitage, J.~P.}
  (\bibinfo{year}{2004}).
\newblock \bibinfo{title}{{Making sense of it all: bacterial chemotaxis.}}
\newblock {\it \bibinfo{journal}{Nat. Rev. Mol. Cell Biol.}\/},  {\it
  \bibinfo{volume}{5}\/}, \bibinfo{pages}{1024--37}.
  \DOIprefix\doi{10.1038/nrm1524}.
\bibitem[{Weiss et~al.(2004)Weiss, Elsner, Kartberg \& Nilsson}]{Weiss2004}
\bibinfo{author}{Weiss, M.}, \bibinfo{author}{Elsner, M.},
  \bibinfo{author}{Kartberg, F.}, \& \bibinfo{author}{Nilsson, T.}
  (\bibinfo{year}{2004}).
\newblock \bibinfo{title}{{Anomalous subdiffusion is a measure for cytoplasmic
  crowding in living cells.}}
\newblock {\it \bibinfo{journal}{Biophys. J.}\/},  {\it
  \bibinfo{volume}{87}\/}, \bibinfo{pages}{3518--24}.
  \DOIprefix\doi{10.1529/biophysj.104.044263}.
\bibitem[{Zhang et~al.(2008)Zhang, Meerschaert \& Baeumer}]{timeLangevin}
\bibinfo{author}{Zhang, Y.}, \bibinfo{author}{Meerschaert, M.~M.}, \&
  \bibinfo{author}{Baeumer, B.} (\bibinfo{year}{2008}).
\newblock \bibinfo{title}{{Particle tracking for time-fractional diffusion}}.
\newblock {\it \bibinfo{journal}{Phys. Rev. E}\/},  {\it
  \bibinfo{volume}{78}\/}, \bibinfo{pages}{36705}.
  \DOIprefix\doi{10.1103/PhysRevE.78.036705}.

\end{thebibliography}

\appendix

\section{Taylor expansions}

For the Taylor expansion of \eqref{eq:markov-master} we use the following
mathematica input:
\begin{verbatim}
(T^+)[x_,h_]:=(1-(S[x+h]-S[x]))q[\[Rho][x+h]]a[\[Rho][x-h]]
(T^-)[x_,h_]:=(1-(S[x-h]-S[x]))q[\[Rho][x-h]]a[\[Rho][x+h]]
A[\[Rho]_,x_,h_]:=(T^+)[x-h,h]\[Rho][x-h]+(T^-)[x+h,h]\[Rho][x+h]
                 -((T^-)[x,h]+(T^+)[x,h])\[Rho][x]
Series[A[\[Rho],x,h],{h,0,2}];
Normal[%]/h^2
Integrate[%,x]
Collect[%,\[Rho]'[x]]
\end{verbatim}

This yields the output 
\begin{align}
3 \rho q(\rho ) \rho ^{\prime }a^{\prime }(\rho )-\rho a(\rho ) \rho
^{\prime }q^{\prime }(\rho )+2 \rho a(\rho ) q(\rho ) S^{\prime }+a(\rho )
q(\rho ) \rho ^{\prime }
\end{align}
from which we read off \eqref{eq:markov-DE}.

For the Taylor expansion of the right-hand side of \eqref{eq:discrete-flux},
we use
\begin{verbatim}
L[x_,h_]:=q[\[Rho][x-h]]a[\[Rho][x+h]] (1/2-(S[x+h]-S[x-h])/4)
R[x_,h_]:=q[\[Rho][x+h]]a[\[Rho][x-h]] (1/2+(S[x+h]-S[x-h])/4)
c[x_,h_]:=1-L[x,h]-R[x,h]
W[i_,x_,h_]:=R[x-h,h]i[x-h]+L[x+h,h]i[x+h]+c[x,h]i[x]
Normal[Series[W[i,x,h]-i[x],{h,0,2}]];
Integrate[%,x];
Expand[%/(h^2/2 a[\[Rho][x]]q[\[Rho][x]])];
Collect[%,i[x]]
\end{verbatim}

which yields the output
\begin{align*}
i(x) \left(\frac{3 \rho ^{\prime }(x) a^{\prime }(\rho (x))}{a(\rho (x))}-%
\frac{\rho ^{\prime }(x) q^{\prime }(\rho (x))}{q(\rho (x))}-2 S^{\prime
}(x)\right)+i^{\prime }(x)
\end{align*}
and we can read off \eqref{transport}.

\section{The method of Characteristics}

We transform the PDE \eqref{structured-dynamics} on the domain $x\in \mathbb{%
R }$, $t>0$, $\tau > 0 $ into an ODE along the characteristics
\begin{align*}
u(s) = (x,\tau,t) - s(0,1,1), \quad s \in [0,\min\{\tau,t\}]
\end{align*}
We write $\xi_\tau $ and $\xi_t $ for the partial derivatives of $%
\xi(\tau,t) $ with respect to the first resp.\ second argument, and find
\begin{align*}
\frac{d}{ds} \xi(u(s)) = -\xi_\tau(u(s)) - \xi_t(u(s)) = [\alpha(u(s)) +
\gamma(u(s))] \xi(u(s))
\end{align*}
where by slight abuse of notation we let $\alpha(x,\tau,t) := \alpha(x,t) $
and $\gamma(x,\tau,t) := \gamma(x,\tau) $. This solves to
\begin{align*}
\xi(u(s)) = C \exp\left( \int_0^s [\alpha(u(y)) + \gamma(u(y))]dy \right)
\end{align*}
and setting $s =0 $ yields the constant $C = \xi(x,\tau,t)$. If $\tau \le t $%
, then
\begin{align*}
\xi(x,0,t-\tau) = \xi(u(\tau)) = \xi(x,\tau,t) \exp\left( \int_0^\tau
[\alpha(x,t-y) + \gamma(x,\tau-y)]dy \right),
\end{align*}
and if $\tau \ge t $, then
\begin{align*}
\xi(x,\tau - t,0) = \xi(u(t)) = \xi(x,\tau,t) \exp\left( \int_0^t
[\alpha(x,t-y) + \gamma(x,\tau-y)]dy \right).
\end{align*}
A change of integration variable together with the definition of $\Phi(x,t) $
and $\Psi(x,\tau) $ then yields \eqref{xi-cases}.

\section{Stationary Distributions}
\label{sec:app-stationary}

The following Mathematica code generates Figure~\ref{fig:no-vf}:
\begin{verbatim}
\[Mu][x_]:=0.7+0.2x 
S[x_]:=2x
\[Alpha][\[Rho]_]:={10^-10,10^-20,10^-30,10^-40,10^-45}
q[\[Rho]_]:=1
a[\[Rho]_]:=1
mass=1;
J[i_,x_]:=-a[\[Rho][x]] q[\[Rho][x]](
           D[i,x]+i D[Log[a[\[Rho][x]]^3/q[\[Rho][x]]]-2 S[x],x])
\[Tau][x_]=10^0;
MLtailL[\[Mu]_,s_,\[Tau]_,x_]:=s^(\[Mu][x]-1)
           /(\[Tau][x]^-\[Mu][x]+s^\[Mu][x]) 
i[x]=\[Rho][x]/MLtailL[\[Mu],\[Alpha][\[Rho][x]],\[Tau],x];
LHS = J[i[x],x];
DElist=Table[Extract[%,i]==0,{i,5}];
Table[NDSolve[{Extract[DElist,i], U'[x]==\[Rho][x],U[0]==0,
      U[1]==mass},{\[Rho],U},{x,0,1}],{i,5}];
nonmarkovsol=\[Rho][x]/.%;
Needs["PlotLegends`"]
Plot[nonmarkovsol,{x,0,1},PlotRange->{{0,1},{0,7}},
     AxesLabel->{x,Subscript[\[Rho], st][x]},
     PlotStyle->{{Black,Dashing[Tiny]},{Black,Dashing[Small]},
     {Black,Dashing[Medium]},{Black,Dashing[Large]},
     {Black,Thick}},PlotLegends->Placed[{
     "log(\[Alpha])=-10","log(\[Alpha])=-20","log(\[Alpha])=-30",
     "log(\[Alpha])=-40","log(\[Alpha])=-45"},{0.8,0.7}]]
\end{verbatim}
note that the Laplace transformed Mittag Leffler function 
\texttt{MLtailL} is chosen according to Eq.(28) in 
\cite{Fedotov2012}:
\begin{align*}
\hat\Psi(x,s) = \frac{s^{\mu(x)-1}}{\tau(x)^{-\mu(x)} + s^{\mu(x)}}
\end{align*}
If Line 4 of the code is replaced by 
\begin{verbatim}
q[\[Rho]_]:=1-0.2 \[Rho]
\end{verbatim}
then the volume filling effect is set to carrying capacity $5 = 1/0.2$, and Figure~\ref{fig:with-vf} results. 
\end{document}